\documentclass[12pt]{article}
\usepackage{amsmath,amsthm, amssymb}
\usepackage{graphicx}
\usepackage{leqno}
\usepackage{hyperref}
\numberwithin{equation}{section}
\begin{document}
\title{\vskip-40pt Wave Function Collapse, Correlating Interactions, and Conservation Laws}
\author{Edward J. Gillis\footnote{email: gillise@provide.net}}

\maketitle

\begin{abstract}

\noindent 
The assumption that wave function collapse is induced by correlating interactions of the kind that constitute measurements leads to a stochastic collapse equation that does not require the introduction of any new physical constants and that is consistent with conservation laws. The collapse operator is based on the interaction (potential) energy, with a variable timing parameter related to the rate at which individual interactions generate the correlations. The approximate localization of physical systems follows from the distance-dependent nature of the interaction potentials. The equation is consistent with strict conservation of momentum and orbital angular momentum, and it is also consistent with energy conservation within the accuracy allowed by the limited forms of energy that can be described within nonrelativistic theory. The possibility of extending the proposal to a fully relativistic version is discussed.

\end{abstract}

\section{Introduction}
\label{intro}

Stochastic, nonlinear modifications of the Schr\"{o}dinger equation have been proposed in order to explain measurement outcomes in terms of fundamental physical processes, and to describe how the apparently linear evolution of systems consisting of a few elementary particles is altered when large numbers of interacting particles become involved. In these accounts wave function collapse and the Born probability rule are consequences of the basic mathematical structure of the theory, rather than ad hoc postulates, seemingly at odds with that structure. This approach has been developed over a number of years in works by various authors\cite{Pearle_1976,Pearle_1979,Gisin_1984,GRW,Diosi_1,Diosi_2,Diosi_3,Gisin_c,GPR,Adler_Brun,Ghirardi_Bassi,Pearle_1,Brody_finite}.

The need to use an equation that is nonlinear in order to describe wave function collapse follows from the fact that projection of a state vector to one of its components is a nonlinear operation.\footnote{A recent work by Mertens, et al.\cite{Mertens_1} has also shown that nonlinearity is required for the derivation of the Born rule.} The need for a nonlinear equation to be stochastic was demonstrated by Gisin who showed that any deterministic nonlinear modification would lead to superluminal signaling\cite{Gisin_c}. Most of the proposals seek to collapse the measured system either to an approximate position state or to an energy eigenstate. Typically, they require the introduction of new physical constants. For example, position-based approaches usually include distance and timing parameters to govern the range and rate of the collapse effects.

 The approach described here originated as an attempt to reconcile relativity with the nonlocal aspects of quantum theory by taking the prohibition of superluminal information transmission  as the fundamental principle that these theories have in common. Consideration of the way in which information is physically instantiated and transmitted led to the hypothesis that the interactions that establish correlations between systems, such as those involved in measurements, are responsible for inducing wave function collapse\cite{Gillis_1}. This hypothesis was formalized in a stochastic collapse equation in \cite{Gillis_2} that used two-particle interaction potentials as the basis for the stochastic operator. In that equation the strength of the collapse effects was determined by the ratio of the interaction energy to the total relativistic energy of the two particles. The timing parameter was assumed to be a new constant that could be chosen to insure that collapse occurred on a time scale consistent with our macroscopic experience. 
 
 The need to introduce such a new constant can be eliminated by tying the collapse rate to the speed with which individual interactions separate the wave function into orthogonal branches in the process of generating correlations. It can be precisely quantified in terms of the rate at which the potential energy changes during the course of the interaction relative to the total energy of the interacting systems. It is defined in such a way that it integrates to a value of order, $1$, over the course of the interaction, and goes to zero when the branching is complete or when the two particles settle into a stationary state.\footnote{This condition holds for those interactions that contribute significantly to the collapse process. For other interactions the integrated value can be much less.} This insures that the collapse equation reduces to the ordinary nonrelativistic Schr\"{o}dinger equation when applied to stationary states or to noninteracting systems.

 Besides eliminating the need to introduce new physical constants there are some other key differences between the equation presented here 
 and most previous proposals. Most approaches have assumed that the correlations between the measured system and the measuring apparatus are established \textit{before} the collapse, and that the collapse occurs on a time scale that is much shorter than that on which ordinary Hamiltonian evolution is effective. In contrast, the process proposed here is assumed to work in an incremental fashion with small transfers of amplitude between distinct branches of the wave function associated with each elementary interaction. Because these incremental stochastic effects are several orders of magnitude smaller than the changes induced by standard Schr\"{o}dinger evolution it works \textit{in parallel with the Hamiltonian} as the correlations are being established. In addition, because the collapse operator is based on the inter-particle potential energy its outcome cannot be readily characterized in terms of an eigenfunction of any single type of operator (such as position or energy). This is due to the fact that the potential energy varies continuously over the course of an entangling interaction, and the integrated value typically varies from one interaction to the next. The collapse basis in each instance is defined by the correlating interactions.\footnote{These branches tend to coincide with  the collapse basis described in the decoherence literature\cite{Zeh,Zurek_ptr,Zurek_Darwin_1,Zurek_Darwin_2}, but the collapse is brought about, primarily, by the interactions between the elementary and other microscopic systems that constitute the measurement apparatus, and, in general, is \textit{not} dependent on the coupling of the apparatus to the environment.}

These distinguishing factors require a different type of proof that the proposed equation is effective in bringing about collapse. As described above, the process works by shifting amplitude between entangled branches depending on the level of interaction potential energy in each branch of the wave function at each stage of the process. So the proof presented here consists, essentially, of an explicit description of the collapse process as a random walk between these developing branches.

  The next section provides a heuristic description of how the collapse process works as illustrated by a simple example. It then lays out the key assumptions that are made and discusses some relevant issues related to entanglement. In Section 3 the collapse operator and equation are defined, and the way in which the relevant parameters are determined is described. These parameters determine both the rate at which collapse occurs and the size of the systems necessary to bring about collapse. Section 4 describes how the wave function evolves in configuration space in measurement situations. It focuses on the way in which the tensor product structure of the Hilbert space and the basis in which to view entanglement relations are defined by the interactions that generate the correlations. Section 5 gives a detailed description of the collapse process in configuration space and a proof that it conforms with the Born probability rule. Section 6 discusses the fact that virtually all elementary systems are entangled to some extent with other systems with which they have interacted, and the implications that this entanglement has for the definition of conserved quantities and our understanding of conservation laws. Based on this discussion Section 7 goes on to demonstrate that the proposed collapse equation is consistent with strict conservation of momentum and angular momentum in individual experiments. Section 8 shows that, within the limits permitted by the somewhat restricted characterization of energy in nonrelativistic theory, consistency with energy conservation is also maintained. Section 9  discusses the possibility of extending the proposal to cover relativistic situations. Section 10 is a summary.

 \section{Motivation, Approach, and Assumptions}
 \label{sec:2} 
 
 The original motivation for assuming that wave function collapse is induced directly by correlating interactions stems from the role that these interactions play in the physical instantiation and transmission of information. Although wave function collapse is a nonlocal effect that \textit{appears} to be at odds with relativity, the fact that it is a probabilistic phenomenon prevents such effects from carrying any information. Thus, it makes sense to focus on the fundamental physical processes involved in enforcing this limit in order to reconcile the ``two fundamental pillars of contemporary theory"\cite{Bell_fund}.

 The emphasis on entangling interactions has important additional benefits.  First, the interactions pick out the basis into which the wave function collapses, and they also help define the tensor product structure of the Hilbert space\cite{Dugic_1,Dugic_2,Bertlmann_2,Bertlmann_1} which is crucial in characterizing the structure of entanglement relations. Second, it allows one to use the strength of the interactions and the duration of the entangling process to eliminate the need to introduce any new physical constants. Finally, it allows the construction of a collapse equation that respects conservation laws \textit{in individual instances}.

Several objections have been raised to the proposed equation based on misunderstandings of how an interaction-induced collapse process works. For this reason it is helpful to begin with a heuristic description such a  process in a very simple measurement situation. Suppose that the wave function of an elementary particle is split into two spatially separated branches, and that a detector is placed in the path of one of these branches. When one of the branches encounters the detector it initiates a sequence or cascade of interactions that correlate the states of the particle to those of the detector. Most collapse proposals assume that these correlations are generated \textit{prior to} the actual collapse.\footnote{In many accounts the \textit{sequence of individual interactions} between the measured system and the the apparatus is idealized by introducing a single effective interaction Hamiltonian. The viewpoint offered here is that this sort of idealization misses the most crucial steps in the collapse process, and creates the impression that the collapse operator must compete with the Hamiltonian, rather than work in parallel with it.} In contrast, the assumption made here is that collapse proceeds incrementally as the entangling interactions are taking place. Specifically, it is assumed that with each such interaction a small amount of amplitude is transferred either into or out of the interacting branch based on some random process. With a sufficient number of interactions eventually all of the amplitude is transferred into either the interacting or the noninteracting branch of the wave function.

The question that naturally arises is what happens if detectors are placed in the paths of both branches. In these situations the process can work just as it does in the case of a single detector, as long as one assumes that the interactions are not perfectly synchronized in both time and strength. Such an assumption is extremely plausible for interaction sequences of sufficient length. 

This intuitive idea must now be formalized in a stochastic collapse equation that covers the wide range of possible measurement situations, and also accounts for collapse in naturally occurring settings. The remainder of this section will set the stage for this by describing the assumptions,  framework, and notation that will be used in carrying out this task.

Although the motivation for this proposal is a desire to reconcile relativity and quantum theory the formulation presented here is nonrelativistic. This simplifies the handling of the nonlocal aspects of wave function collapse and makes it possible to analyze the collapse process in configuration space. The configuration space picture is also helpful in explaining how the process respects conservation laws. The nonrelativistic formulation also entails that the kinetic and interaction energies are much less than the total relativistic energy. 

The total system is assumed to consist of a very large number of subsystems. Much of the discussion will focus on elementary particles or other very small subsystems, but we must also allow for the possibility of elementary systems organizing themselves into much larger, even macroscopic, subsystems that can interact with other subsystems as a unit. Individual subsystems will be labeled with subscripts, $j,k,l,...$ Space coordinates will be labeled as $x_j, y_j, z_j$, and their vector representation will be represented as $\mathbf{w}_\textbf{j}$. 

Interactions will be modeled by conservative (hence, distance-dependent), two-particle potentials, $\mathbf{\hat{ V}_{jk}}(\mathbf{w}_j,\mathbf{w}_k) 
\, = \, \mathbf{\hat{ V}_{jk}}(|\mathbf{w}_j \, - \,\mathbf{w}_k|) $, that fall off with increasing distance. The term, 'particle', is to be interpreted as applying to any subsystem interacting as a unit. Since I will be assessing the effect of the collapse operator on conservation laws it is also necessary to rule out external potentials. All systems, even macroscopic ones, are considered to be quantum systems. There is no classical boundary.

By basing the collapse operators on the potentials, which are smooth functions of distance that decrease in magnitude as distance increases, we can insure that that the collapse results in the approximate localization of subsystems. The special status of the position basis follows from the assumption that collapse is tied to distance-dependent interactions. The close connection between the interaction basis and the position basis will also facilitate the description of the collapse process and the maintenance of conservation laws in configuration space.

It is also assumed that it is possible to measure any (reasonable) observable. Since measurements consist of sets of interactions, and since interactions are distance dependent, this implies that measurements typically begin with the spatial separation of the wave function of the target system into branches characterized by distinct values of the quantity associated with the observable being measured. This leads to the splitting of the wave function of the total system into well separated regions of configuration space. The random walk described in Section 5 takes place between these regions.

 \section{Stochastic Collapse Equation}
 \label{sec:3}

Like a number of previous proposals, the collapse equation described here is based on the Wiener process, also known as Brownian motion. The derivative of this process is white noise, which is usually indicated in the literature as $dW$, $dB$, or $d\xi$. A fairly standard form for these equations is:\footnote{Some recent proposals employ colored noise rather than white noise, and incorporate the stochastic processes into a classical noise field. See \cite{Adler_Bassi,Adler_Vinante,Bassi_Ferialdi}.}  
\begin{equation}\label{3x1}   
d\psi\, \,   = \,  (-i/\hbar)\mathbf{\hat{H}} \, \, \psi\, dt \,  +\, 
\Big{ [} \sum_k \sqrt{\gamma}(\mathbf{\hat{ L}_k} - \langle \, \mathbf{\hat{ L}_k} \, \rangle) d\xi_k 
- \, \frac{1}{2}\sum_k\gamma(\mathbf{\hat{ L}_k} - \langle \, \mathbf{\hat{ L}_k} \, \rangle)^2  dt \Big{ ]}\psi .  
\end{equation} 
The first term on the right represents the ordinary Schr\"{o}dinger evolution, governed by the Hamiltonian, $\mathbf{\hat{H}}$. The $\mathbf{\hat{ L}_k}$ are Lindblad operators which are almost always taken to be self-adjoint; hence, they correspond to observables. This schema allows for multiple independent Wiener processes, which can be real or complex. $\langle \, \mathbf{\hat{ L}_k} \, \rangle$ is the expectation value of $\mathbf{\hat{ L}_k} $ in the state, $\psi$, $\langle \psi | \mathbf{\hat{ L}_k}| \psi \rangle$, and its presence makes clear the nonlinearity of the equation. The parameter,  $\gamma$, determines the effectiveness of the process in bringing about collapse, and, in most proposals, incorporates new physical constants. The stochastic processes, $\xi_k$, are assumed to have zero mean, and the differentials, $d\xi_k$, obey the It$\hat{o}$ stochastic calculus rules: 
\begin{equation}\label{3x2}   
d\xi^{*}_{j} d\xi_k = dt \delta_{jk}, \;\;\; dt d\xi_k = 0. 
\end{equation}

These conditions on the differentials determine the units,  
$ d\xi_k \,\sim \, (dt)^{\frac{1}{2}}$, and they are responsible for the factor, $dt$, in the final term on the right. The middle term, $\sum_k\sqrt{\gamma}(\mathbf{\hat{ L}_k} - \langle \, \mathbf{\hat{ L}_k} \,  \rangle)  d\xi_k $, is primarily responsible for bringing about the collapse of the state vector. In most other proposed equations it tends to concentrate the state on eigenstates of the observables, 
$\mathbf{\hat{ L}_k}$\cite{Adler_Brun}. (The approach taken here differs in that the set of possible outcomes is determined by which interactions are most relevant in a particular situation.) The nonunitary nature of the operator described by middle term necessitates an adjustment to insure that the resulting vector has unit norm. This is provided by the third term,    
$- \, \frac{1}{2}\sum_k\gamma(\mathbf{\hat{ L}_k} - \langle \, \mathbf{\hat{ L}_k} \, \rangle)^2  dt $.

The collapse equation proposed here will use just a single collapse operator, $ \hat{\mathcal{V}}, $ and a single Wiener process, $\xi$. In order to implement the hypothesis that wave function collapse is induced by those interactions that generate entanglement the stochastic operator will be based on conservative, two-particle interaction potentials, 
$\mathbf{\hat{ V}_{jk}}(\textbf{w}_j,\textbf{w}_k)$, where $\textbf{w}_j$ and $\textbf{w}_k$ indicate the coordinates of systems $j$ and $k$ in configuration space. The requirement that the potential functions are conservative implies that they are functions only of the separation between the systems: $\mathbf{\hat{ V}_{jk}}(\mathbf{w}_j,\mathbf{w}_k) \, = \, \mathbf{\hat{ V}_{jk}}(|\mathbf{w}_j \, - \,\mathbf{w}_k|) $.

The emphasis here is on \textit{correlating} interactions. So we want to multiply the potential by some measure of the effectiveness of the interaction in establishing a correlation between the systems involved. Since this depends on the extent to which the interaction changes the states of the systems involved, the expression,  
\newline 
$\mathbf{\hat{ V}_{jk}} \, - \, \langle \, \mathbf{\hat{ V}_{jk}} \, \rangle$, will be divided by the sum of the \textit{effective masses}, $ (m_j+m_k) $, of the interacting systems. For the present purpose the most straightforward way to determine an effective mass is to aggregate particles in bound states such as atoms, molecules, and other complex structures into single systems with a net charge and a total mass.

To maintain overall consistency in the dimension of the stochastic operator it is necessary to transform the effective mass in the denominator into an energy by multiplying by the square of some velocity. The only nonarbitrary choice is $c$, the speed of light. As will be shown, multiplying $\mathbf{\hat{ V}_{jk}} \, - \, \langle \, \mathbf{\hat{ V}_{jk}} \, \rangle$ by $ 1 \, / \, (m_j+m_k)c^2 $ sets the size of the amplitude shifts associated with individual interactions to a very reasonable value, and insures that the  stochastic effect is a minor perturbation relative to the changes induced by standard Schr\"{o}dinger evolution. It also anticipates the eventual extension  of this proposal to a relativistic version.

The other critical piece of the stochastic operator is the rate parameter, $\gamma$. As stated earlier the single parameter, $\gamma$, will be replaced by multiple variable rate parameters,  $\gamma_{jk}$, 
each of which is associated with the interaction between system, $j$, and system, $k$. These are tied to the speed with which individual interactions separate the wave function into orthogonal branches in the process of generating correlations. They are effective \textit{only during the brief period in which the correlation is being established}, and they tend to zero afterward. They are defined below. 

The individual terms, 
$ [\sqrt{\gamma_{jk}}(\mathbf{\hat{ V}_{jk}} - \langle \mathbf{\hat{ V}_{jk}} \rangle)] \; / \; [(m_j+m_k)c^2] $, will be abbreviated as $\hat{\mathcal{V}}_{jk}.$ The single collapse operator, $ \hat{\mathcal{V}}, $ is then defined as: 
$\sum_{j<k} \hat{\mathcal{V}}_{jk}. $ The proposed collapse equation can be represented as either: 
\begin{equation}\label{3x3}    
d\psi \,   = \,  (-i/\hbar)\mathbf{\hat{H}} \, \psi \, dt \,  +\, 
\sum_{j < k} \hat{\mathcal{V}}_{jk}   \, \, \psi\, d\xi(t)  
- \, \frac{1}{2}  (\sum_{j < k} \hat{\mathcal{V}}_{jk}) \,  
(\sum_{m < n} \hat{\mathcal{V}}_{mn} ) \, \psi \, dt ;  
\end{equation}
or, more compactly, as: 
\begin{equation}\label{3x4}    
d\psi \,   = \,  (-i/\hbar)\mathbf{\hat{H}} \, \psi \, dt \,  +\, 
 \hat{\mathcal{V}} \, \psi\, d\xi(t)  
- \, \frac{1}{2}  \hat{\mathcal{V}}^2 \,   \psi \, dt.  
\end{equation}

It is important to emphasize that the amplitude transfer brought about by each $\hat{\mathcal{V}}_{jk}$ applies across the entire configuration space and not only to the systems, $j$ and $k$. Other systems are affected by the transfer to the extent that they are entangled with either $j$ or $k$. Because the potentials, $\mathbf{\hat{ V}_{jk}}$, are distance-dependent the amplitude transfers occur between localized regions of configuration space.

To complete the collapse proposal it is necessary to specify the rate parameters,  $\gamma_{jk}$. These will be defined in terms of the rate at which the potential energy is changing during the interaction between system, $j$, and system, $k$. They vanish for stationary states since the interaction potential is constant in these situations. They also tend to zero when the two systems become widely separated.

The rate of change of potential energy is:
\begin{equation}\label{3x5} 
 \frac{d}{dt} \, \langle  \, \psi | \, \mathbf{\hat{ V}_{jk}}  \, | \psi, \rangle \; 
 = \; \frac{-i}{\hbar} \Big{(} \, \langle \, \psi | \,
 \Big{[}  \mathbf{\hat{ H}_{jk}}, \mathbf{\hat{ V}_{jk}} \Big{]} \, | \psi \rangle \, \Big{)},
\end{equation}
where $ \Big{[}  \mathbf{\hat{ H}_{jk}}, \mathbf{\hat{ V}_{jk}} \Big{]} $ is the commutator of $  \mathbf{\hat{ H}_{jk}} $ and $ \mathbf{\hat{ V}_{jk}},  $
and the nonrelativistic Hamiltonian is: 
\begin{equation}\label{3x6}  
\mathbf{\hat{ H}_{jk}} \;\; = \;\; 
-\frac{\hbar^2}{2m} (\mathbf{\nabla_j}^2  \, + \,  \mathbf{\nabla_k}^2)  \;
+  \; \mathbf{\hat{ V}_{jk}}. 
\end{equation}
This yields:
\begin{equation}\label{3x7} 
(-\frac{i \hbar}{2m}) \, \langle \, \psi | \, (\mathbf{\nabla_j}^2   \mathbf{\hat{ V}_{jk}} \, + \, \mathbf{\nabla_k}^2   \mathbf{\hat{ V}_{jk}}    \; + \;   
2 ( \mathbf{\nabla_j}  \mathbf{\hat{ V}_{jk}} 
\mathbf{\nabla_j}  
\, + \, \mathbf{\nabla_k} \mathbf{\hat{ V}_{jk}} \mathbf{\nabla_k}) \, | \psi \rangle \,   .
\end{equation}
The nonrelativistic limit on energy entails a minimum for the term, $|\mathbf{w}_j \, - \,\mathbf{w}_k| $. This, in turn, implies that the integration over $\psi$ eliminates the terms, 
$\mathbf{\nabla_j}^2   \mathbf{\hat{ V}_{jk}}$ and 
$\mathbf{\nabla_k}^2   \mathbf{\hat{ V}_{jk}}.$
The result is:
\begin{equation}\label{3x8} 
|| \, \int \, \psi^* \, \Big{[} (\mathbf{\hat{\nabla}_j}\mathbf{\hat{ V}_{jk}} \,  \cdot \,  ( \frac{-i \hbar}{m} ) \, \mathbf{\nabla_j} \psi) \, + \,
(\mathbf{\hat{\nabla}_k}\mathbf{\hat{ V}_{jk}} \,  \cdot \,  ( \frac{-i \hbar}{m} ) \, \mathbf{\nabla_k} \psi) 
\Big{]} \, ||. 
\end{equation}
This expression includes the norm ($||...||$) in order to insure a positive value. (All integrals are over the entire configuration space unless otherwise noted.)

Some further refinement of this formula is required. As described earlier  typical measurement situations begin with the branching of the wave function of an elementary or other very small system, $j$, into spatially separated regions. Only one of these branches encounters system, $k$, and interacts with it. It is essential for the derivation of the Born probability rule that the rate parameter, $\gamma_{jk}$, be independent of the amplitude of the interacting branch. For this reason we want to pick out just the interacting portion of the wave function. This also allows one to define $\gamma_{jk}$ in such a way that it integrates to a value of order, $1$, over the duration of the interaction.

The interacting portion of the wave function can be picked out by using   a weighting factor, $ \frac{ \mathbf{\hat{ V}_{jk}} \,} { \langle \, \mathbf{\hat{ V}_{jk}}  \, \rangle }  $, reflecting the extent to which each segment of the wave function is involved in the interaction. To insure that $\gamma_{jk}$ has the dimensions of inverse time and that it integrates to a value of order, $1$, the rate of change of potential energy will be divided by the total nonrelativistic energy of the interacting systems. Since the kinetic energy of the center of mass of the interacting systems adds an arbitrary offset to the total energy it is necessary to eliminate it by specifying the energy in the center-of-mass frame. The center-of-mass velocity can be identified by integrating the relevant momentum operators over the interacting portion of the wave function and dividing by the the combined mass:
\begin{equation}\label{3x9} 
\vec{\mathbf{u }} \; = \; 
\int \,  \Big{(}\frac{ \mathbf{\hat{ V}_{jk}} \,}
{ \langle \, \mathbf{\hat{ V}_{jk}}  \, \rangle } \,\Big{)} \, \psi^* \,  \,  ( \frac{-i \hbar}{mj+m_k} ) \, [ \, \mathbf{\nabla_j} \psi \, + \, \mathbf{\nabla_k} \psi \, ].
\end{equation}  
 The wave function in the center-of-mass rest frame can then be represented as: 
\begin{equation}\label{3xa}
\psi_{cm} \; = \; \Big{(}\frac{ \mathbf{\hat{ V}_{jk}} \,}
{ \langle \, \mathbf{\hat{ V}_{jk}}  \, \rangle } \,\Big{)} \, \psi \, \, e^{ i [ - \, \vec{\mathbf{u}} \cdot  \vec{\mathbf{w}}_j  \frac{m_j}{\hbar}  \, - \,  m_j \mathbf{u}^2 \frac{t}{2 \hbar} \, - \, \vec{\mathbf{u}} \cdot  \vec{\mathbf{w}}_k  \frac{m_k}{\hbar}  \, - \,  m_k \mathbf{u}^2 \frac{t}{2 \hbar}]}.
\end{equation}
With this expression the relevant measure for the total energy is:
 \begin{equation}\label{3xb} 
 \langle \, \psi_{cm} \, | \,  \Big{(} \, \hat{\mathbf{H}}_{jk} \, - \, E_0 \,  \Big{)}         \,  | \, \psi_{cm} \, \rangle,  
 \end{equation}
where $E_0$ is the ground state energy. (It should be noted that the terms, $ \frac{ \mathbf{\hat{ V}_{jk}} \,} { \langle \, \mathbf{\hat{ V}_{jk}}  \, \rangle }  $, $ \vec{\mathbf{u }} $, and, hence, $\psi_{cm}$ can vary over the duration of the interaction.) 
Putting together \ref{3x8} through \ref{3xb} we arrive at the definition for 
 $\gamma_{jk}$ : 
\begin{equation}\label{3xc} 
\gamma_{jk} \; \equiv \; 
\Big{[} \, 
|| \, \int \, \psi^* \, \Big{[} (\mathbf{\hat{\nabla}_j}\mathbf{\hat{ V}_{jk}} \,  \cdot \,  ( \frac{-i \hbar}{m} ) \, \mathbf{\nabla_j} \psi) \, + \,
(\mathbf{\hat{\nabla}_k}\mathbf{\hat{ V}_{jk}} \,  \cdot \,  ( \frac{-i \hbar}{m} ) \, \mathbf{\nabla_k} \psi) 
\Big{]} \, ||\, \Big{]}  \;  \Big{/}   \; \Big{[} \, 
 \langle \, \psi_{cm} \, | \,  \Big{(} \, \hat{\mathbf{H}}_{jk} \, - \, E_0 \,  \Big{)}         \,  | \, \psi_{cm} \, \rangle  
\, \Big{]}.
\end{equation}

Our primary interest is in the integrated value of $\gamma_{jk}$ for those pairs of systems that interact most strongly since these interactions are chiefly responsible for inducing the collapse. It should be clear that in these cases the integrated value has a maximum of order, $1$, since, in the center-of-mass frame one expects an approximately complete changeover from potential to kinetic energy (or vice versa). In a transition from a free state to a free state the value would be about $2$. Transitions to bound states would be about $1$. (Radiative effects are neglected since this is a nonrelativistic formulation.) Since $ \int \, \gamma_{jk} \, dt  \; \sim \; 1$ it is also clear that 
$ \parallel \, \int \, \sqrt{\gamma_{jk}} \, d\xi(t) \parallel \; \sim \; 1. $ 
In more weakly interacting cases the change in potential energy is much smaller than the average kinetic energy over the course of the interaction, and so the integrated value of $\gamma_{jk}$ and the effect on the amplitude are much less.

For the cases of primary interest, the fact that $\gamma_{jk}$ integrates to about $1$ allows us to estimate the size of the amplitude shift associated with each interaction. The shift will be roughly equal to the ratio of interaction energy to total relativistic energy. The upper limit for this value can be estimated by considering electrostatic potentials and electron-electron interactions since these have the lowest mass of any particles in a nonrelativistic theory. The upper limit on the interaction energy must be large enough to accommodate typical measurement situations, while respecting the nonrelativistic constraint, $\mathbf{\hat{V}}_{jk} \,  \ll  \, mc^2$.  The ratio of potential energy to relativistic energy for an electron in the ground state of a hydrogen atom is equal to the square of the fine structure constant, 
$ \alpha^2  \, \approx \,  5.33*10^{-5}$. This number provides a convenient reference, but it is not quite large enough to cover all measurements of interest.\footnote{In photomultipliers the electrons involved in the measurement process are accelerated up to several hundred electron volts. Since one would not expect all of this energy to be concentrated in a single electron-electron interaction this limit should be sufficient.} Therefore, it is reasonable to consider interaction energy ratios of roughly one order of magnitude above this reference,  
$\mathbf{\hat{V}}_{jk} \; / \; [(m_j+m_k)c^2]   \sim  \, 10 \alpha^2 \,  \sim  \, 5*10^{-4}$. Energy ratios much above this value can no longer be described with simple distance-dependent potentials, and relativistic corrections would need to be taken into account.

It is also of interest to estimate the the ratio of the change in the wave function due to the stochastic operator relative to that induced by the Hamiltonian over the course of the interaction. This ratio can be expressed as:
\begin{equation}\label{3xd}    
\Big{[} \int \, \hat{\mathcal{V}}_{jk}   \, \, \psi\, d\xi(t)  \Big{]}   \; / \;  
\Big{[} \int \, \frac{\mathbf{\hat{V}}_{jk}}{\hbar} \, \psi \, \delta t \Big{]}
\end{equation}
Since $\hat{\mathcal{V}}_{jk}$ is defined as $[\sqrt{\gamma_{jk}}(\mathbf{\hat{ V}_{jk}} - \langle \mathbf{\hat{ V}_{jk}} \rangle)] \; / \; [(m_j+m_k)c^2] $, and $ \parallel \, \int \, \sqrt{\gamma_{jk}} \, d\xi(t) \parallel \; \sim \; 1, $ this ratio is approximately  
$  [ 1 \, / \, (m_e+m_e)c^2] *[\hbar \, / \, \delta t].  $  The time interval, $\delta t$, is inversely proportional to the $\frac{3}{2}$ power of the interaction energy, and the mass of the electron, $m_e$, has again been used to determine an upper limit.

The interaction energy limit described above can be used to estimate the relevant time interval, $\delta t$, by considering the time it takes for the potential energy to change from $0.1$ of its maximum value to its maximum value (or vice versa). The maximum speed of an electron at the energy limit is about $7*10^6 m/s$ and the distance over which the change in   
$ \mathbf{\hat{ V}_{jk}} $ takes place is about equal to the Bohr radius, $\approx \, 5* 10^{-11}m$. So $\delta t$ is about $10^{-17}$ seconds. The resulting ratio is:
\begin{equation}\label{3xe}    
[ 1 \, / \, (m_j+m_k)c^2] *[\hbar \, / \, \delta t] \; \approx \;  
[1 \, / \, 10^{-13}] *[ 10^{-34} \, / \,10^{-17} ] \; = \; 10^{-4}.
\end{equation}

The smallness of this ratio shows that the largest stochastic effect from a single interaction is a fairly small nonlinear perturbation on the Hamiltonian. Both of the ratios derived here play key roles in the discussion in Section 5 which shows how \ref{3x3} brings about the collapse of the wave function. In preparation for that demonstration the next section deals with some issues involved in the evolution of the wave function and the generation of entanglement relations in measurement situations.

\section{Evolution and Entanglement in Configuration Space}
\label{sec:4} 

As just shown the effects of the stochastic operator from a single interaction are minor perturbations on the standard Schr\"{o}dinger evolution. So, to explain how \ref{3x3} induces complete collapse it is helpful to first review the way in which the wave function evolves in  typical measurement situations under the influence of the Hamiltonian. This background is assumed, either explicitly or implicitly, by every serious attempt to understand what happens during quantum measurements.

The interactions that constitute measurements are distance dependent. For this reason measurements intended to determine a particular quantity typically begin by splitting the subject wave function into branches characterized by different values of that quantity. For example, to measure momentum one might send the subject particles through a magnetic field that bends the trajectories in various directions. To measure spin along a particular axis one can also use a magnetic field to separate up and down states. In some cases, of course, the separation occurs simply as a result of the evolution of the  wave function.

Some or all of the branches then proceed into devices in which the subject particle can initiate a set of interactions that result in a final  state of the device that is macroscopically distinct from its initial state. The arrangement of interactions that carry out this task must be designed in such a way that an elementary or other microscopic system is capable of completely altering the state of a macroscopic system. 

In many accounts the detailed interactions involved in correlating the state of an elementary system with that of the measurement apparatus are ignored. Many stochastic collapse proposals reduce this process to a single step described by a very simple Hamiltonian. In a similar fashion, most of the literature on decoherence assumes that it is the interactions between the apparatus and the ``environment" that are most relevant in explaining the appearance of particular outcomes.

In contrast, in this proposal the individual interactions that take place between the measured system and the measurement apparatus, along with those \textit{within the apparatus}, play the crucial role in bringing about the collapse of the wave function.\footnote{This is not meant to rule out the possibility that subsequent interactions between the apparatus and environment might help to finish the process.}

The critical role that individual, localized interactions play in the development of distinct branches of the wave function is evident in Mott's analysis of $\alpha$-ray tracks\cite{Mott}. By examining the evolution of the wave function in configuration space he showed how these interactions lead to correlations between the positions of the atoms that are ionized by the passage of the $\alpha$-particle. These correlations are responsible for the well defined tracks that are seen in these experiments. Since Mott's paper numerous studies have shown how the establishment of correlations between elementary systems leads to the suppression of interference between various segments of the wave function. The reason for this suppression can be easily understood by considering the description of the evolving system in configuration space. The interactions lead to the separation of the wave function into essentially disjoint regions. This point has been elaborated in several discussions of the branching process in configuration space in connection with the de Broglie-Bohm theory.\cite{Bell_dBR_Bhm,Bell_imp_p_w,Bohm_Hiley,Norsen_PW,Romano} 

Because the representation of the wave function in configuration space exhibits the clear separation of the branches, it provides a convenient way to understand how the measurement basis is selected. It also enables a detailed description of the collapse process according to \ref{3x3}. It is, therefore, worth examining this representation in more detail.

Each point in configuration space represents a classical configuration of the total system. The regions in which the wave function differs significantly from zero contain the configurations that are consistent with the prior evolution of the system, in particular, with the interaction history. This history determines how conserved quantities have been exchanged during interactions, and the extent to which the states of various subsystems are correlated. 

The correlations play a key role in defining the entanglement structure of the wave function, but entanglement also depends on how the total system is decomposed into subsystems. In other words, it depends on the choice of a coordinate system in configuration space. For the purpose of analyzing the evolution of the wave function in measurement situations the natural choice for a coordinate system is the one used to describe distance-dependent interaction potentials, and is often referred to simply as the position basis. This corresponds to a decomposition of the total system into \textit{interacting} subsystems. That is the decomposition that will be used here. (Alternate decompositions that decouple evolution equations by eliminating any reference to interactions between systems will be reviewed briefly in the next section. These decompositions eliminate entanglement by changing the tensor product structure of the Hilbert space.)

The separate localized segments of the wave function reflect the entanglement structure defined by the position basis. The branches of systems that have interacted and exchanged conserved quantities are represented in the same region of configuration space and their complementary branches are in a different region.

The entanglement relations can be examined in a little more detail by considering the effect of an interaction between two elementary systems. Suppose that system, $j$, is the subject of the measurement, and that a branch of this system interacts with system, $k$, in a region around $(\mathbf{w}_j,\mathbf{w}_k)$. The systems exchange momentum and energy during the interaction, and, as a result, their trajectories are altered in a coordinated manner reflecting the entanglement that now exists between them. The components of the systems that interacted around $(\mathbf{w}_j,\mathbf{w}_k)$ are now linked and 
will be represented in a region of configuration space around $(\mathbf{w}'_j,\mathbf{w}'_k)$. When either of these components interacts with another system, $l$, it extends the chain of entanglement relations. In this manner further measurement-like interactions build a large structure of correlated components that define a macroscopically distinct state of the total system, represented in a disjoint, localized region in configuration space. 

Consider the density, $\psi^* \psi$, corresponding to the system wave function in just this entanglement region. Under ordinary Schr\"{o}dinger evolution, governed only by the Hamiltonian, it will have some integrated value, say, $\mu^*\mu$. To demonstrate that \ref{3x3} is effective in bringing about the collapse of the wave function it must be shown that, as a result of the stochastic action associated with each of the elementary correlating  interactions that constitute the measurement, the integrated value, $\mu^* \mu$, is altered to either $1$ or to $0$. This is done in the next section.

 \section{Collapse of the Wave Function}
 \label{sec:5}

To begin the demonstration it is necessary to calculate the change in the quantity, $\psi^* \psi$, induced by each correlating interaction according to \ref{3x3}. We first restrict \ref{3x3} to a single interaction: 
\begin{equation}\label{5x1}    
d\psi \,   = \,  (-i/\hbar)\mathbf{\hat{H}} \, \psi \, dt \,  +\, 
\hat{\mathcal{V}}_{jk}   \, \, \psi\, d\xi(t)  
- \, \frac{1}{2}  \hat{\mathcal{V}}_{jk}^2 \,   \psi \, dt. 
\end{equation}
The adjoint equation is:
\begin{equation}\label{5x2}    
d\psi^* \,   = \,  (+i/\hbar)\mathbf{\hat{H}} \, \psi^* \, dt \,  +\, 
\hat{\mathcal{V}}_{jk}   \, \, \psi^* \, d\xi^*(t)  
- \, \frac{1}{2}  \hat{\mathcal{V}}_{jk}^2 \,   \psi^* \, dt. 
\end{equation}
The change in the product,  $\psi^*\psi$, is: 
\begin{equation}\label{5x3} 
d(\psi^*\psi) \; = \; (\psi^* + d\psi^*)(\psi + d\psi) \, - \, (\psi^*\psi) \; = \;   d\psi^* \psi \, + \psi^*  d\psi \,  +  d\psi^* d\psi. 
\end{equation} 
To calculate the change we must take into account the It$\hat{o}$ calculus rules regarding the way in which differentials are treated. In ordinary calculus a term like $d\psi^* d\psi $ that is quadratic in differentials would be discarded. But the 
Wiener process, $ \xi(t) $,  varies as $ \sqrt{t}$, since it is analogous to a random walk. It is this relationship that is responsible for the It$\hat{o}$  rule stated in \ref{3x2}, ($d\xi^{*} d\xi = dt $), and it also implies that the final term in \ref{5x3},  $d\psi^{*} d\psi $, must be retained.   

Plugging \ref{5x1} and \ref{5x2} into \ref{5x3} yields:   
\begin{equation}\label{5x4}  
\begin{array}{ll}
d(\psi^*\psi) \; = \;  
 \frac{i}{\hbar} \Big{[} \, (\mathbf{\hat{H}} \, \psi^*) \psi 
 \, - \, \psi^* (\mathbf{\hat{H}} \psi) \, \Big{]} \, dt  
& \\  
- \psi^* (\frac{1}{2}\hat{\mathcal{V}}_{jk} 
\hat{\mathcal{V}}_{jk})   \, \psi  \, dt \,  
- \psi^* (\frac{1}{2}\hat{\mathcal{V}}_{jk} 
\hat{\mathcal{V}}_{jk} )  \, \psi  \, dt \,  
+ \, \psi^* \, \hat{\mathcal{V}}_{jk}   \hat{\mathcal{V}}_{jk} 
\psi  \, d\xi^*d\xi \,   
& \\    
\; \;\; \;   \; \;\; \; \; \;  \; \;\; \;  \;\; \;\; + \; \; 
\psi^* \,    \hat{\mathcal{V}}_{jk}   \psi    \, d\xi^* \, +\,  \psi^* \,    \hat{\mathcal{V}}_{jk}  \psi  \, d\xi. 
\end{array}
\end{equation}
Since $d\xi^{*} d\xi = dt $ we get:   
\begin{equation}\label{5x5}     
\;\;\;   d (\psi^* \psi) \; = \; 
- \, \mathbf{\nabla}  \cdot   \Big{[}  
( \frac{i \hbar}{2 m} )
\,  ( \, \psi  \mathbf{\nabla}\psi^* \, - \,  \psi^*  \mathbf{\nabla}\psi) \Big{]} \, dt  \;  + \; 
(\psi^* \,    \hat{\mathcal{V}}_{jk} \, \psi)  \, (d\xi^* 
\, + \,  d\xi).  
\end{equation}  
The first term on the right is just the negative of the divergence of the probability current, which represents the change in $\psi^* \psi$ under ordinary Schr\"{o}dinger evolution.

The Schr\"{o}dinger term and the stochastic term, 
$(\psi^* \,    \hat{\mathcal{V}}_{jk} \, \psi)  \, (d\xi^* 
\, + \,  d\xi),$ operate in essentially complementary and independent ways in measurement-like situations. The change induced in $\psi^* \psi$ by the Hamiltonian is relatively large and local, while the change resulting from the stochastic term is relatively small and nonlocal. As described in the previous section, an effective measurement process is one in which the individual interactions are arranged so that they clearly separate the wave function into disjoint regions of configuration space. The separation between the various branches is several orders of magnitude greater than the effective range of the measurement interactions.

The change induced by the Hamiltonian on the interacting portion of the wave function is, by definition, confined to the interaction region. Thus, the action of the Hamiltonian on each branch is effectively independent of its action on other branches, and since it is unitary it is also independent of the amplitude of each branch.

The effect of the stochastic operator \textit{is to transfer amplitude between the branch(es) undergoing interaction and the disjoint, noninteracting branches}. The change induced by the individual transfers in the value of $\psi^*\psi$ integrated across each branch is given by \ref{5x3}: 
$(\psi^* \, \hat{\mathcal{V}}_{jk} \, \psi) 
\, (d\xi^* \, + \,  d\xi)$. The direction of the transfer depends on the sign of the product of 
$\hat{\mathcal{V}}_{jk} \, (d\xi^* \, + \,  d\xi)$, and its magnitude was discussed in Section 3. Because the term, $\sqrt{\gamma_{jk}} \, d\xi$, integrates to approximately $1$, the size of each transfer depends on the ratio of the interaction energy to the total relativistic energy of the systems involved, along with the value of $\psi^* \psi$. Thus, with each individual correlating interaction there is a transfer of amplitude either into or out of the interacting portion of the wave function. Again, it is important to note that the amplitude transfer applies to the \textit{total system} - not only to the interacting subsystems. Thus, the entire branch of the wave function in which the interaction takes place has its amplitude altered by the stochastic action.\footnote{Although this point seems obvious it has been missed by several critics.} 
Since the wave function is assumed to be normalized and equations of the form, \ref{3x1}, are norm-preserving the integrated value of $\psi^* \psi$ is limited by $0$ and $1$. Therefore, with a sufficient number of correlating interactions this integrated value must approach one of these limits. Measurement processes typically involve enough individual interactions of adequate strength to insure complete collapse. More detailed quantitative estimates will be provided below.

Despite the straightforward manner in which \ref{3x3} brings about collapse there have been several claims that it is incapable of doing so. These are dealt with here.

It has been objected that the collapse process could be frustrated if measurement-like interactions are occurring in several branches simultaneously. However, in situations in which there is a sufficient number of interactions the probability of this is vanishingly small, as can be seen by considering the numerical estimates from Section 3. The estimate for the maximum ratio of interaction energy to total relativistic energy in nonrelativistic situations was about $5*10^{-4}$. The amplitude transfer process has the character of a random walk, with the step size determined by this ratio. The endpoints of the random walk are $0$ and $1$ for the integrated values of $\psi^* \psi$. The typical number of steps required to reach an endpoint is the square of the inverse of the step size. Given the maximum value for the step size just quoted, this means that the smallest number of individual interactions to complete the process would be in excess of $10^6$. The white noise that helps to determine the direction of the steps varies continuously. This means that to frustrate the collapse the interactions in disjoint branches of the wave function would have to be almost perfectly synchronized in both time and interaction strength over a walk of this length. An individual interaction with this strength has a duration of about $10^{-17}$ to $10^{-16}$ seconds and a range of less than $10^{-10}$ meters. The branches are separated by at least a few orders of magnitude greater than the effective range of the interactions. Variations in propagation delays and the complexity of propagation patterns make it nearly impossible to achieve the synchronization that would be needed. Interactions of lower strength would have longer durations and expanded ranges, but the required number of steps increases with the square of the inverse interaction strength. So the necessary synchronization remains a virtual impossibility.

Another objection that has been raised is that in the cases in which the collapse process is effective it would necessarily reduce the wave function to a single point, implying an infinite increase in energy. This objection is based on a misunderstanding of the way in which \ref{3x3} differs from most collapse equations. Most proposals are designed to collapse the wave function to an eigenfunction of one particular observable, for example, either position or total energy. Since the stochastic operator in \ref{3x3} is based on distance-dependent interaction potentials it has been alleged that it must reduce the wave function to an eigenfunction of potential energy, that is, a single point. As discussed previously, the effect of the operator relative to the Hamiltonian is very small, \textit{and the duration of its action is extremely short}. Moreover, this effect is smeared over the interacting portion of the wave function by the ordinary Hamiltonian evolution. There is simply no way that collapse to a single point can occur.\footnote{The extent to which the wave function is reshaped by the collapse operator is examined in section 8.}

One additional criticism of the current proposal is that because it conserves momentum (as will be shown in Section 7) it cannot be effective in bringing about collapse. The argument is that when momentum is conserved the center of mass evolves unitarily. This feature is alleged to prevent collapse since collapse equations are necessarily nonunitary.

This objection fails because, as long as the dynamics governing the system insures conservation of momentum, the motion of the center of mass \textit{is completely decoupled from} the evolution of the individual components that make up the system. Equations that redefine the the system into independently evolving subsystems (such as in simple two-body problems) alter the tensor-product structure of the Hilbert space by eliminating interactions between the components, and changing the entanglement structure. This change in entanglement structure has been discussed by  Jekn\'{i}c-Dug\'{i}c, Arsenijev\'{i}c, and Dug\'{i}c, and by Thirring, Bertlmann, K\"{o}hler, and Narnhofer\cite{Dugic_1,Dugic_2,Bertlmann_2,Bertlmann_1}. The authors of \cite{Bertlmann_2} note that most discussions of entanglement implicitly assume a decomposition of the total system into \textit{interacting} subsystems:
\begin{quote}
	``Thus, it's only the interaction, which we consider to determine the density
	matrix, or the measurement set-up, which fixes the factorization."
\end{quote}
The center of mass, regarded as an individual subsystem, is completely disentangled from the remainder of the system. It is not subject to any interactions or external potentials, and so it evolves freely and unitarily. But this implies \textit{nothing} about whether the rest of the system also evolves in a unitary fashion.

This concludes the proof that \ref{3x3} is effective in inducing wave function collapse in typical measurement processes. Note that the reference to the way in which the wave function evolves in these processes has been used in order to facilitate this demonstration. However, it should be clear that generally similar branching processes also occur in natural settings, outside the laboratory. In fact, they are ubiquitous. Thus, one would expect wave function collapse, induced in this manner, to occur all the time.

What remains is to show that the probability of collapse to a particular branch of the wave function complies with the Born rule. Consistency with the Born Rule follows essentially from the general form of \ref{3x1}. This can be shown in detail for \ref{3x3} as follows. 

Assuming an adequate number of sufficiently strong correlating interactions the process described by \ref{3x3} can be characterized as a one-dimensional random walk with variable step size that ends with an integrated value of $\psi^* \psi$ over the interacting component of the the wave function equal to either $0$ or $1$. If one assumes that this integrated value at some stage of the process is equal to $\mu^* \mu$ then to establish the Born rule we need to show that the probability that the end value will be $1$ is $\mu^* \mu$.

We begin with a lemma that applies to simple one-dimensional random walks. Label one end point as $0$ and the other as $1$, and let the distance between them be divided into $n$ equal intervals. An elementary theorem states that from a point, $m$, the probability of the walk terminating at $1$ is $\frac{m}{n}$. An analogous result holds if steps of variable size are allowed. From a point, $p$, with $ 0 \, \leq \, p \, \leq \, 1$ the probability of reaching $1$ is $p$. To see this, let $\delta$ be a variable increment that is less than or equal to the distance to the nearest end point: 
$ 0 \, \leq \, p - \delta \, \leq \, p \, \leq \, p + \delta \, \leq \, 1$. Designate the probability of reaching $1$ as $Pr(p)$. We have 
$Pr(0) \, = \, 0$, $Pr(1) \, = \, 1$, and 
$ Pr(p) \, = \, \frac{1}{2}[ Pr(p-\delta)] \, + \, 
\frac{1}{2}[ Pr(p+\delta)]$. Since the last condition holds for all values of $p$ and $\delta$, $Pr(p)$ must be linear. Given the boundary conditions it follows that $ Pr(p) \, = \, p.$ Since a Wiener process is the scaling limit of a one-dimensional random walk, this result holds whether the walk is regarded as a discrete or continuous process. 

In the collapse process $p$ is replaced by the value of $\psi^* \psi$ integrated over the interacting component, 
$ \mu^* \mu$. The corresponding value for the complementary (noninteracting) component is $\nu^* \nu \, =  1 \, - \, \mu^* \mu$. What must be shown is that the step size at each stage of the collapse process is less than or equal to both $\nu^* \nu $ and $ \mu^* \mu$.

The stochastic change in $\psi^* \psi$ from a single correlating interaction is given by \ref{5x5}: 
$(\psi^* \,    \hat{\mathcal{V}}_{jk} \, \psi)  \, (d\xi^* 
\, + \,  d\xi)$. The individual terms, $\hat{\mathcal{V}}_{jk}$, were defined in Section 3 as 
$ [\sqrt{\gamma_{jk}}(\mathbf{\hat{ V}_{jk}} - \langle \mathbf{\hat{ V}_{jk}} \rangle)] \; / \; [(m_j+m_k)c^2] $. We first calculate the value of $(\mathbf{\hat{ V}_{jk}} - \langle \mathbf{\hat{ V}_{jk}} \rangle) $ across the interacting and the noninteracting component of $\psi$. The value of 
$\mathbf{\hat{ V}_{jk}}$ across the noninteracting component is essentially zero. Although the value varies somewhat across the interacting component, it is substantially different from that of the noninteracting portion. Designate the interacting component as $I$, and the orthogonal component as $O$. The average value of $\mathbf{\hat{ V}_{jk}}$ over $I$ is 
$ \overline{\mathcal{ V}^I_{jk}} \; \equiv \;  [ \, \langle \, I \, |V_{jk}| \, I \, | \rangle \,] \; / \; [ \, \langle \, I \, | \,  I \, \rangle \, ]$. The average of $\mathbf{\hat{ V}_{jk}}$ over the entire wave function is then: $ \langle \mathbf{\hat{ V}_{jk}} \rangle \; = \, \mu^*\mu \, 
\overline{\mathcal{ V}^I_{jk}} $. The average value of $(\mathbf{\hat{ V}_{jk}} - \langle \mathbf{\hat{ V}_{jk}} \rangle) $ across the interacting  component is 
$ \overline{\mathcal{ V}^I_{jk}} \, (1 \, -  \mu^*\mu \, ) 
 \; = \; \nu^* \nu \, \overline{\mathcal{ V}^I_{jk}} \, $. The corresponding value averaged over the orthogonal component is 
$ - \, \mu^* \mu \, \overline{\mathcal{ V}^I_{jk}} \, $. Using these averaged values, the expression, 
$(\psi^* \,    \hat{\mathcal{V}}_{jk} \, \psi)  $, can be written as: \newline 
$ [ \,  \langle \, I \, | \, \nu^* \nu \, \overline{\mathcal{ V}^I_{jk}} \, | I \, \rangle \; - \;  \langle \, O \, | 
\mu^* \mu \,  \overline{\mathcal{ V}^I_{jk}} \, | O \, \rangle \, ] \;
/ \; [(m_j+m_k)c^2]. $ Since $\overline{\mathcal{ V}^I_{jk}} $ is constant over the wave function this can be represented as: 
\newline 
$ \overline{\mathcal{ V}^I_{jk}} \, [ \nu^* \nu \,  \langle \, I \, |  \, I \, \rangle \; - \; \mu^* \mu \, \langle \, O \,  
  | O \, \rangle ] \;
/ \; [(m_j+m_k)c^2]. $

We have $  \langle \, I \, |  \, I \, \rangle \; = \, \mu^*\mu $ and  $  \langle \, O \, |  \, O \, \rangle \; = \, \nu^*\nu $. Therefore, it is clear that the stochastic operator associated with each individual interaction multiplies each of the two orthogonal components by equal and opposite amounts: 
\newline 
$ \pm \; \mu^* \mu \, \nu^* \nu \, \overline{\mathcal{ V}^I_{jk}} \;
/ \; [(m_j+m_k)c^2]. $
Since this relationship holds for all pairs of systems, $j,k$, if multiple interactions are taking place at the same time we simply sum over them. The full stochastic effect on the various branches at each time is given by: 
\begin{equation}\label{5x8}
 \pm \; (\mu^* \mu \, \nu^* \nu) \,  \mathbf{ \sum_{jk}} \, (\psi^* \; \overline{\mathcal{ V}^I_{jk}} \; / \; [(m_j+m_k)c^2] \,  \sqrt{\gamma_{jk}} \; \psi)  \, (d\xi^*(t) \, + \,  d\xi(t)). 
\end{equation} 

The estimate for the maximum ratio of $\overline{\mathcal{ V}^I_{jk}} \; / \; [(m_j+m_k)c^2] $ from Section 3 was $5*10^{-4}$. The terms, $ \sqrt{\gamma_{jk}} $, are tied to the rate at which the interaction proceeds and, when conjoined with the terms, $d\xi^*(t) \, + \,  d\xi(t))$, are designed to integrate to a value of about $1$ over the course of the interaction. The stochastic process, $ \xi(t)$, has a variance of $t_2 - t_1$ over the period from $t_1$ to $t_2$. Therefore, to insure that the norm of the sum,  
$ \mathbf{ \sum_{jk}} \, (\psi^* \; \overline{\mathcal{ V}^I_{jk}} \; / \; [(m_j+m_k)c^2] \,  \sqrt{\gamma_{jk}} \; \psi)  \, (d\xi^*(t) \, + \,  d\xi(t))$, is less than or equal to $1$, we need only to take time increments, $dt$, that are sufficiently small. Since $\mu^* \mu$ and $\nu^* \nu$ are both less than or equal to $1$ this will guarantee that the total changes in the branches of $\psi^* \psi$ given by \ref{5x8} are less than or equal to $1$. Thus, \ref{3x3} reproduces the Born rule.

It should be clear that the number of correlating interactions in a typical measurement instrument during a measurement process is sufficient to bring about nearly complete collapse. Instruments consist of well in excess of $10^{23}$ particles and a substantial fraction of these are involved in the chain or cascade of interactions that result in macroscopic differences in the state of the device. With an average step size of $s$ the number of steps required to complete a random walk between the point $0$ and $1$ is on the order of $ 1 / s^2$. The estimates of Section 3 showed that the maximum step size is about $5*10^{-4}$. Measurement interactions could be a few orders of magnitude less than this and still be effective. Measurements are often initiated with individual interaction energies comparable to those of visible photons, and these would give interaction energy ratios in the range of $10^{-6}$ to $10^{-5}$. However, subsequent interaction energies in the process are frequently enhanced by the instrument. It seems reasonable to take the range of $10^{-7}$ to $10^{-6}$ as a lower limit for interaction energy ratios in most measurement situations. 

The other key factor in determining the step size is the product, $\mu^* \mu  \nu^* \nu$. As $\mu^* \mu$ approaches either $0.99$ or $0.01$ this term shrinks to a magnitude of about $0.01$, reducing the step size by a couple of orders of magnitude. With the estimates from the previous paragraph this would set an upper limit on the number of steps required at about $10^{16}$ to $10^{17}$. It is important to note that, as the measurement proceeds, the great majority of these interactions would be occurring in parallel - not sequentially. Weaker interactions would take about $10^{-11}$ seconds to complete, but since so many would be taking place at the same time the overall completion time for the process would be some small fraction of a second.

The way in which \ref{3x3} insures consistency with conservation laws in individual situations will be reviewed in the next few sections.

 \section{Conserved Quantities Are Shared Across Entangled Branches}
 \label{sec:6}

 A proper understanding of conserved quantities in quantum theory must start with a recognition of the fact that \textit{all} systems are quantum systems. The idealized simplifications that are used to analyze particular types of situations and the classical boundaries that are placed around these situations must be seen as just pragmatic approximations. The use of such approximations is dictated by the impossibility of tracking all the interactions that couple a ``system" to its environment. But the fact that we cannot do a detailed calculation of the effect of all of these interactions does not prevent us from drawing general conclusions about their effect by applying some fundamental results of quantum theory.

 One such key result is that entanglement is a \textit{generic} effect of the interaction between systems. This has been demonstrated by Gemmer and Mahler\cite{Gemmer_Mahler} and by Durt\cite{Durt_a,Durt_1}. An important implication of this result is that there is almost always some small amount of entanglement between an elementary system that has been ``prepared" in an (approximate) eigenstate of some observable and the preparation apparatus.  
 
 A photon interacting with a beam-splitter provides a simple example of such entanglement between a system and apparatus. (This example has been discussed in unpublished works by this author\cite{Gillis_3} and by Marletto and Vedral\cite{Marletto_Vedral}.) During the interaction the reflected branch of the photon exchanges momentum with the beam-splitter, and alters its state by a tiny amount. The extremely narrow wave function representing the beam-splitter in position space implies an extremely large spread in the momentum representation of the beam-splitter's state. Thus the exchange of momentum with the photon branch alters that state by a very tiny amount. The magnitude of the inner product between the slightly altered mirror state and the pre-interaction state of the apparatus can be represented as $ |1-\delta| $, where $ \delta \ll 1 $, but $ \delta >  0$. This nonzero $\delta$ implies some (very small) amount of entanglement between the photon and the beam-splitter. If, for the sake of simplicity, one assumes that the photon branches have equal amplitude, the entanglement, in terms of $\delta$, can be approximated as $ (\delta/2) | 1 - \log(\delta/2)  | $\cite{Gillis_3}. It is important to note that momentum is strictly conserved within each branch of the entangled wave function.

 If the photon is subsequently detected in one of the branches then, because it is entangled with the beam-splitter, the resulting collapse of the wave function encompasses not only the photon but also the correlated state of the beam-splitter. No matter which branch is detected the total momentum of the photon, beam-splitter and measurement apparatus is the same after the detection as it was before the initial photon-mirror interaction.    
 
 Because of the generic nature of the relationship between interaction and entanglement this conservation of the relevant quantities among ``system", preparation apparatus, and measurement instrument holds in virtually all situations. The fact that these quantities are exchanged and shared across branches of entangled systems through interactions implies that definite values of conserved quantities can be assigned to these entangled branches - not just to elementary subsystems that are assumed to be in factorizable eigenstates of the relevant observable.   
 
 We have no trouble in assigning definite values of conserved quantities to  branches of entangled systems when those systems consist of a small number of elementary particles such as singlet states, correlated photons generated during parametric down conversion, or the kinds of states typically considered in discussions of quantum computation. But, we lose sight of the fact that well-defined quantities can be distributed across branches when those branches include somewhat larger subsystems.

 The aim of this and the next two sections is to show that the proposed collapse equation (\ref{3x3})is consistent with the conservation laws of standard, unitary quantum theory in individual measurement situations in regard to momentum, angular momentum, and energy. Given this, the foregoing considerations about entanglement require a careful examination of exactly what those laws imply. A quantity, $\mathbf{q}$, is conserved if the observable, $\mathbf{\hat{Q}}$, with which it is associated commutes with the evolution operator, $\mathbf{\hat{H}}$, for arbitrary states, $| \psi \rangle$:  
 $\mathbf{\hat{H}} \mathbf{\hat{Q}} | \psi \rangle
 \; = \;  \mathbf{\hat{Q}} \mathbf{\hat{H}} | \psi \rangle \, $. When this condition holds the time derivative of $\mathbf{q}$, integrated over the state, $| \psi \rangle$, is zero: 
 $ (d/dt) \langle \, \psi \, | \mathbf{\hat{Q}} | \psi \rangle \; = \; \frac{i}{h} \,  \langle \, \psi \, | \mathbf{\hat{H}} \mathbf{\hat{Q}} \, - \,  \mathbf{\hat{Q}} \mathbf{\hat{H}} | \psi \rangle \, $. 
 
 Note that these laws are \textit{not} restricted to situations in which the the state, $| \psi \rangle$, is (assumed to be) in an eigenstate of the operator, $\mathbf{\hat{Q}}$. The argument presented earlier in this section implies that the systems under consideration are almost never in exact eigenstates. To impose such a restriction would render the conservation laws essentially vacuous.

 The question of whether the conservation laws hold in a particular situation depends not only on the structure of equation \ref{3x3}, but also on the initial conditions that are assumed. The claim being made here is only that \ref{3x3} \textit{is consistent with} these laws. What will be shown in the next two sections is that the proposed equation guarantees that, within each entangled branch,\footnote{The tensor product structure and the basis in which entangled branches are defined is assumed to be that which is picked out by the interactions. This basis coincides with the position basis in configuration as described in Section 4.} the relevant quantities, $\mathbf{q}$, are conserved, when the branch is normalized. In order to establish overall consistency it must be assumed that the normalized quantities in different (affected) branches are equal. The remainder of this section will explain why this assumption is reasonable for a wide range of initial conditions.

 A key background assumption of the proposed collapse equation  is that the ``system" under consideration consists of a very large number of subsystems that have been interacting over an extended period. For such a system the properties of any elementary subsystem are determined essentially by its interactions with other subsystems, with the most recent interactions having the greatest effect. Branching processes are initiated by these \textit{conservative} interactions. Consider again the example of the photon and beam-splitter. It is clear that the interaction between them does not introduce any difference in momentum between the reflected and transmitted branch. The momentum change in the reflected branch of the photon is offset by a momentum change in one of the (slightly different) beam-splitter states. The total momentum in each (normalized) branch is the same after the interaction as it was prior to it. The same kind of analysis can be applied to any conservative interaction, involving any conserved quantity. Therefore, after an extended period of interaction one would expect that the total value of these quantities would be essentially the same in all of the branches.

 This argument can be extended to cover varying configurations within each branch. As the wave function evolves under the influence of the Hamiltonian it continually recombines the configurations. Thus, what distinguishes the configurations at different points is the distribution of various quantities among the subsystems - not the total value of those quantities computed across all subsystems. In other words, after a sufficient number of interactions along with ordinary wave propagation, the total value of a conserved quantity in a particular configuration is very close to the average value of that quantity calculated across the branch. Therefore, in what follows it will be assumed that the total value of conserved quantities is the same at each point in configuration space where the wave function has nonzero amplitude. 
 
 With this assumption in the next two sections it will be shown that \ref{3x3} is consistent with strict conservation of momentum, angular momentum, and energy in individual measurements. The demonstrations for momentum and angular momentum are fairly straightforward, and they hold exactly. These cases are dealt with in Section 7. Because of nonrelativistic limitations the situation with energy is somewhat more complicated. It is dealt with separately in Section 8.

\section{Momentum and Angular Momentum}
\label{sec:7} 

In this section I will show that the stochastic operator defined in section 3 maintains the \textit{normalized value} of momentum and angular momentum at each point in configuration space. In other words the change in these quantities induced by the collapse equation is directly proportional to the change that is induced in the squared amplitude at that point:
\begin{equation}\label{7x1}
d(\psi^*\mathbf{\hat{Q}}\psi) \, / \,  (\psi^*\mathbf{\hat{Q}}\psi)  \; = \; d(\psi^* \, \psi) \, /  \, (\psi^* \, \psi).
\end{equation}
Together with the assumption described at the end of Section 6 this implies that these quantities are strictly conserved under evolution governed by \ref{3x3}.

The change in $d(\psi^* \, \psi) $ was displayed in equation \ref{5x5}:
\newline 
$   d (\psi^* \psi) \; = \; 
- \, \mathbf{\nabla}  \cdot   \Big{[}  
( \frac{i \hbar}{2 m} )
\,  ( \, \psi  \mathbf{\nabla}\psi^* \, - \,  \psi^*  \mathbf{\nabla}\psi) \Big{]} \, dt  \;  + \; 
(\psi^* \,    \hat{\mathcal{V}}_{jk} \, \psi)  \, (d\xi^* 
\, + \,  d\xi).  $
The steps leading up to \ref{5x5} can be repeated to compute $d(\psi^*\mathbf{\hat{Q}}\psi) $:     
\begin{equation}\label{7x2}   
\begin{array}{ll} 
d (\psi^* \, \mathbf{\hat{Q}} \,\psi) \; = \; 
\frac{i}{\hbar} \Big{[} \, (\mathbf{\hat{H}} \, \mathbf{\hat{Q}} \,\psi^*) \psi 
\, - \, \psi^* (\mathbf{\hat{Q}} \,\mathbf{\hat{H}} \psi) \, \Big{]} \, dt  
\; 
& \\    
\; \;\; \;   \; \;\; \; \; \;  \; \;\; \;   \; \;\; \;   \;\; \;\;  \; \; 
 -\psi^* (\frac{1}{2}  \hat{\mathcal{V}}
\hat{\mathcal{V}}) \, \,\mathbf{\hat{Q}} \, \psi \, dt \; - \;
\psi^* \, \hat{\mathbf{Q}} \, (\frac{1}{2}\hat{\mathcal{V}}
\hat{\mathcal{V}}) \, \psi \, dt \,   

+ \; \psi^* \, \hat{\mathcal{V}}  \hat{\mathbf{Q}}  \hat{\mathcal{V}} \,  \psi   \, d\xi^*d\xi
\; 
& \\    
\; \;\; \;   \; \;\; \; \; \;  \; \;\; \;   \; \;\; \;   \;\; \;\; + \; \; 
\psi^* \,  \hat{\mathcal{V}}  \hat{\mathbf{Q}} \, \psi  \, d\xi^* \; + \;\psi^* \,  \hat{\mathbf{Q}}   
\hat{\mathcal{V}} \,  \psi  \, d\xi 
\; 
& \\    
\; \;\; \;   \; \;\; \; \; \;  \; \;\; \;   \; \;\;         
\; = \;   \;  
-\frac{1}{2}\psi^* [(\hat{\mathcal{V}})^2 \mathbf{\hat{Q}}  \, + \,
\hat{\mathbf{Q}}  (\hat{\mathcal{V}})^2 ]\psi \, dt    \;   + \; \psi^*( \hat{\mathcal{V}}  \hat{\mathbf{Q}}  \hat{\mathcal{V}} \, )\psi   \, dt \; \;   
& \\  
\; \;\; \;   \; \;\; \; \; \;  \; \;\; \;   \; \;\; \;   \;\;       
\; \; + \; \;  \psi^* \,  \hat{\mathcal{V}}  \hat{\mathbf{Q}} \, \psi  \, d\xi^* \; + \;\psi^* \,  \hat{\mathbf{Q}}   
\hat{\mathcal{V}} \, \psi  \, d\xi.   
\end{array}
\end{equation}

The first term in \ref{7x2} represents the change due to ordinary Schr\"{o}dinger evolution. By assumption $\mathbf{\hat{H}}$ commutes with $\mathbf{\hat{Q}}$, but commutativity of the operators only implies that 
$\mathbf{\hat{H}} \mathbf{\hat{Q}}  \, = \, 
\mathbf{\hat{Q}}\mathbf{\hat{H}} $, when they are integrated over the entire wave function. We need a somewhat stronger condition. We need to show that the change in $(\psi^*\mathbf{\hat{Q}}\psi)$ matches the change in  
$(\psi^* \, \psi)$ at each point in configuration space:
$ d(\psi^*\mathbf{\hat{Q}}\psi)  \; \sim \; d(\psi^* \, \psi).$ For the quantities under consideration here this can be seen by observing that the probability current, $   
( \frac{i \hbar}{2 m} ) \, ( \, \psi  \mathbf{\nabla}\psi^* \, - \,  \psi^*  \mathbf{\nabla}\psi) \, $, is simply (except for the factor, $m$, in the denominator) the expression for momentum in a slightly disguised form. In a manner of speaking the quantities, momentum and angular momentum, ``ride along" with the changes in $(\psi^* \, \psi)$ on the probability current. Thus, the changes due to Schr\"{o}dinger evolution in both $(\psi^*\mathbf{\hat{Q}}\psi)$ and $(\psi^* \, \psi)$, decouple from those induced by the collapse operator. 

Therefore, we are left with the task of showing that the change in $(\psi^*\mathbf{\hat{Q}}\psi)$  resulting from the collapse operator, 
\newline
$ -\frac{1}{2}\psi^* [(\hat{\mathcal{V}})^2 \mathbf{\hat{Q}}  \, + \, \hat{\mathbf{Q}}  (\hat{\mathcal{V}})^2 ]\psi \, dt    \;   + \; \psi^*( \hat{\mathcal{V}}  \hat{\mathbf{Q}}  \hat{\mathcal{V}} \, )\psi   \, dt \; + \; \;  \psi^* \,  \hat{\mathcal{V}}  \hat{\mathbf{Q}} \, \psi  \, d\xi^* \; + \;\psi^* \,  \hat{\mathbf{Q}} \hat{\mathcal{V}} \,\psi \, d\xi$, is proportional to $(\psi^* \,    \hat{\mathcal{V}} \, \psi)  \, (d\xi^*  \, + \,  d\xi)$. This can be done by showing that
\begin{equation}\label{7x3} 
\hat{\mathbf{Q}}   \hat{\mathcal{V}} \, \psi 
\; = \; \hat{\mathcal{V}} \, \hat{\mathbf{Q}} \psi 
\end{equation} 
at each point in configuration space.

The only spatial dependence in the collapse operator, $\hat{\mathcal{V}} $, is in the terms representing the conservative interaction potentials, 
$\mathbf{\hat{ V}_{jk}}(\mathbf{w}_j \, - \, \mathbf{w}_k) $. Since this implies that 
$ \mathbf{\hat{\nabla}_j} \, \mathbf{\hat{ V}_{jk}}  \; = \;  -\mathbf{\hat{\nabla}_k} \, \mathbf{\hat{ V}_{jk}}  $ it is quite simple to show that the relation, \ref{7x3}, holds for the two operators in which we are interested here, momentum and orbital angular momentum.

The momentum operator can be expanded as: 
\begin{equation}\label{7x4} 
\mathbf{\hat{P}} \, \equiv \,  -i \hbar \sum_i \,\mathbf{\hat{\nabla}_i}. 
\end{equation} 
Because the action of $\mathbf{\hat{\nabla}_j}  $ cancels that of $\mathbf{\hat{\nabla}_k}  $ when acting on $ \mathbf{\hat{ V}_{jk}},  $ and given the fact that 
$ \mathbf{\hat{\nabla}_i} \, \mathbf{\hat{ V}_{jk}}  \; = \; 0 $ for $i \, \neq \, j, \; \; i  \, \neq \, k $ it is clear that the operator, $\mathbf{\hat{P}} $, simply passes through $\hat{\mathcal{V}} $, and acts only on the wave function, $\psi$. So the normalized value of momentum is conserved at every point in configuration space.

The orbital angular momentum operator can be represented as: 
\begin{equation}\label{7x5}
 \mathbf{\hat{L}} = -i \hbar \sum_i \, \mathbf{w_i}   \mathbf{\times}      \mathbf{\hat{\nabla}_i}.
\end{equation}
Explicit expansions for the x-component for systems, $j$, and $k$, are as follows: 
\begin{equation}\label{7x6} 
\mathbf{\hat{L}_x}(\mathbf{w_j}) \;  = \;  -i\hbar(y_j\partial{\mathbf{\hat{V}_{jk}}}/\partial{z_j}  -  z_j\partial{\mathbf{\hat{V}_{jk}}}/\partial{y_j}). 
\end{equation} 
\begin{equation}\label{7x7} 
\mathbf{\hat{L}_x}(\mathbf{w_k}) \;  = \;  -i\hbar (y_k\partial{\mathbf{\hat{V}_{jk}}}/\partial{z_k}  -  z_k\partial{\mathbf{\hat{V}_{jk}}}/\partial{y_k}). 
\end{equation}  
 The action of $ \mathbf{\hat{L}_x} $ on each $\mathbf{\hat{V}_{jk}} $ is given by: 
 \begin{equation}\label{7x8} 
 \begin{array}{ll} 
[ \mathbf{\hat{L}_x}(\mathbf{w_j}) \, + \, \mathbf{\hat{L}_x}(\mathbf{w_k}) ] 
\mathbf{\hat{V}_{jk}} 
 \;  = \; 
 -i\hbar[  (y_j \, + \, y_k) (\partial{\mathbf{\hat{V}_{jk}}}/\partial{z_j} 
 \, + \, \partial{\mathbf{\hat{V}_{jk}}}/\partial{z_k} )
 & \\     \; \;\; \;  \;\; \;   \;\; \;  \;\; \;   \; \; 
 \; \;\; \; \; \;\; \; \; \;  \; \;\; \;  \; \;\; \; \;\; \;\;  \; \;\; \;
 \; - \;   
 (z_j \, + \, z_k) (   \partial{\mathbf{\hat{V}_{jk}}}/\partial{y_j}  
 \, + \, \partial{\mathbf{\hat{V}_{jk}}}/\partial{y_k} )]. 
 \end{array} 
 \end{equation}   
Since we have both:   
  \begin{equation}\label{7x9} 
  (\partial{\mathbf{\hat{V}_{jk}}}/\partial{z_j}   \, + \, \partial{\mathbf{\hat{V}_{jk}}}/\partial{z_k} ) \; = \; 0; 
     \; \;\; \;  \;\; 
  (\partial{\mathbf{\hat{V}_{jk}}}/\partial{y_j}   \, + \, \partial{\mathbf{\hat{V}_{jk}}}/\partial{y_k} ) \; = \; 0,
  \end{equation}  
it is clear that  $ \mathbf{\hat{L}} $, like $ \mathbf{\hat{P}} $, effectively operates only on the wave function, $\psi$, and not on $\hat{\mathcal{V}} $. So, \ref{3x3} also strictly conserves orbital angular momentum at a point.

This completes the demonstration that \ref{3x3} is consistent with strict conservation of momentum and angular momentum in individual measurement situations. The issue of energy conservation is dealt with in the next section.

\section{Energy and Nonrelativistic Limitations}
\label{sec:8}

Conventional nonrelativistic quantum theory accurately characterizes the status of energy conservation for stationary states and in situations with freely evolving noninteracting particles. In these states there is no radiation, and even though the theory does not represent relativistic corrections to kinetic energy associated with mass increase, there is no breach of conservation since the correction terms do not change.

The proposed collapse equation agrees with conventional theory in these cases since it reduces to the Schr\"{o}dinger equation when there is no exchange of energy between systems. This is obvious when there is no interaction, and the collapse operator also goes to zero in stationary states since it includes the terms, $\gamma_{jk}$, which are based on the rate of change of potential energy, which is zero in these situations.

In all other situations when there is an exchange of energy between systems there is some inaccuracy in the conventional theory's prediction of perfect conservation of kinetic-plus-potential energy. This is due to the fact that it lacks the means to represent some forms of energy. These include radiation, relativistic corrections to the kinetic energy formula, antiparticles, and interactions that cannot be fully characterized with scalar potentials.

The collapse operator in \ref{3x3} does induce a change in energy at each point in configuration space that is proportional to the change in $\psi^* \psi$ just as in the cases of momentum and angular momentum analyzed in the previous section. This accounts for the primary change in energy that occurs in measurement situations,\footnote{There are also energy changes induced by the measurement apparatus, but these do not pose any problem for energy conservation.} and this change is consistent with strict energy conservation.

However, in addition to correctly accounting for the change in energy in the measured subsystem the proposed collapse equation does predict some small deviations from strict conservation of energy. This is due to the fact that it is based on the conventional nonrelativistic theory and that it is designed to act in exactly the situations in which the conventional theory fails to account for all of the energy changes. What will be shown here is that these deviations originate from exactly the same types of interactions that lead to small errors in the predictions of the conventional theory. The fact that the deviations predicted by the collapse proposal are qualitatively and quantitatively similar to the inaccuracies of the conventional theory strongly suggests that they should be regarded as a consequence of the nonrelativistic formulation of both theories, rather than as an indication of a real problem with energy conservation.

The change induced in a particular quantity by the collapse process was given in equation \ref{7x2}. In parallel with the calculations of the previous section the effects of the collapse operator on energy can be evaluated by considering the relationship between   
$ \mathbf{\hat{H}} \, \hat{\mathcal{V}}_{jk} $ and $ \hat{\mathcal{V}}_{jk} \,\mathbf{\hat{H}}.$ Since it is clear that, at each point in configuration space, 
$\mathbf{\hat{V}} \hat{\mathcal{V}}  
\; = \; \hat{\mathcal{V}}  \mathbf{\hat{V}}, $ the only deviations from perfect conservation that can arise are those involving the kinetic energy terms. Because the kinetic energy operator involves the second derivative the expansion of $-\frac{\hbar^2}{2m} \mathbf{\nabla}^2 \, \hat{\mathcal{V}}_{jk} $ is more complicated than the earlier calculations:
\begin{equation}\label{8x1} 
\begin{array}{ll} 
(-\frac{\hbar^2}{2m} \mathbf{\nabla}^2) \, \hat{\mathcal{V}}_{jk}  
\; \; =   \; \;
\hat{\mathcal{V}}_{jk}
(-\frac{\hbar^2}{2m} (\mathbf{\nabla_j}^2  \, + \,  \mathbf{\nabla_k}^2) ) \; 
&  \\  \; \; \; \; \; \; \; \; \; \; \; \;  +  \; \;
(-\frac{\hbar^2}{2m}) \, \{ \, (\mathbf{\nabla_j}^2   \hat{\mathcal{V}}_{jk} \, + \, \mathbf{\nabla_k}^2   \hat{\mathcal{V}}_{jk}    \; + \;   
2 ( \mathbf{\nabla_j}  \hat{\mathcal{V}}_{jk} \, \cdot \, 
\mathbf{\nabla_j}  
\, + \, \mathbf{\nabla_k} \hat{\mathcal{V}}_{jk} \, \cdot \,  \mathbf{\nabla_k})\}  .
\end{array}
\end{equation}
As in the cases of momentum and angular momentum the changes in energy induced by the Hamiltonian and the collapse operator decouple. The collapse-induced changes can be computed by
substituting $\mathbf{\hat{H}}$ for $\mathbf{\hat{Q}}$ in the relevant part of \ref{7x2}, using \ref{8x1}, and expanding: 
\begin{equation}\label{8x2} 
\begin{array}{ll}
d (\psi^* \, \mathbf{\hat{Q}} \,\psi)_C \; =  \; \; 
\psi^* \, \hat{\mathcal{V}}  \mathbf{\hat{H}} \, \psi \, \, (d\xi^* \, + \, d\xi) \;  
& \\
&  \\  - \; \;
(\frac{\hbar^2}{2m}) \,  \psi^* \,  \{ \sum_{j<k} \, [  (\mathbf{\nabla_j}^2  \hat{\mathcal{V}}_{jk} \, + \, \mathbf{\nabla_k}^2  \hat{\mathcal{V}}_{jk}) \psi   \; + \;   
2 (\mathbf{\nabla_j}\hat{\mathcal{V}}_{jk}  \, \cdot \, 
\mathbf{\nabla_j}\psi  
\, + \, \mathbf{\nabla_k}\hat{\mathcal{V}}_{jk} \, \cdot \,  \mathbf{\nabla_k}\psi)]\} \, d\xi  
& \\
& \\   + \; \;     
(\frac{\hbar^2}{2m})  \psi^*   \psi  \,   
(\mathbf{\nabla} \hat{\mathcal{V}} \cdot  \mathbf{\nabla} \hat{\mathcal{V}} )   \, dt. 
\end{array}
\end{equation}

The first line expresses the desired proportionality between $ d (\psi^* \, \mathbf{\hat{Q}} \,\psi)_C$ and $d (\psi^* \, \psi)_C$, and mirrors the relationships for momentum and angular momentum derived in the previous section. If it were not for the deviations shown on the second and third line this would imply that strict energy conservation is maintained within the nonrelativistic theory. If it can be shown that the problematic terms displayed on the second and third lines are artifacts of the nonrelativistic formulation, and that they do not represent real deviations then the case will have been made that wave function collapse is consistent with strict conservation of energy.\footnote{As argued in Section 6, the common presumption that the apparent narrowing of the wave function associated with collapse \textit{must} lead to violations of energy conservation stems from the failure to recognize the entanglement of the measured system with systems with which it has previously interacted and the fact that these systems are also involved in the collapse.}

Unless a very high level of precision is required relativistic corrections to the nonrelativistic expressions become relevant only at fairly high energies. The magnitudes of the inaccuracies in the predictions of the conventional theory are well below the level of precision that is usually required. It will be shown here that the apparent deviations implied by the collapse proposal are of similar magnitudes. There are substantial differences between the deviations represented by the terms on the second and third line of \ref{8x2}, and so they will be dealt with separately.

Consider an interaction between system $j$ and system $k$, and focus on just the effect on system $j$. Using the definition of $\hat{\mathcal{V}}_{jk}$ from Section 3 the relevant part of the second line can be expanded as:
\begin{equation}\label{8x3}
 (-\frac{\hbar^2}{2m}) \mathbf{\big{[}} \psi^* \, \psi    (\mathbf{\nabla_j}^2 \, \mathbf{\hat{ V}_{jk}}  ) 
\; + \;      
2 \psi^*(\mathbf{\nabla_j} \mathbf{\hat{ V}_{jk}}  \, \cdot \, 
\mathbf{\nabla_j}\psi) \mathbf{\big{]}}
(\frac{1}{ (m_j+m_k) c^2)} ) 
\, (\sqrt{\gamma_{jk} }\, d\xi) .  
\end{equation}
When there is a change in kinetic energy some of that change is unaccounted for in the conventional theory due to relativistic change in mass. (There is also some radiation, although it is several orders of magnitude less than the kinetic energy discrepancies.) So let us compare the deviations expressed in \ref{8x3} to the rate at which potential energy is converted to kinetic energy according to the Schr\"{o}dinger equation: 
\begin{equation}\label{8x4}
 (\frac{i \hbar}{2m}) \mathbf{\big{[}} \psi^* \, \psi    (\mathbf{\nabla_j}^2 \, \mathbf{\hat{ V}_{jk}}  ) 
\; + \;      
2 \psi^*(\mathbf{\nabla_j} \mathbf{\hat{ V}_{jk}}  \, \cdot  \,
\mathbf{\nabla_j}\psi)  \mathbf{\big{]}} . 
\end{equation}
The expressions in the square brackets in \ref{8x3} and \ref{8x4} are identical. This illustrates the fact that the deviations implied by \ref{3x3} arise in exactly the same situations in which the conventional theory fails to account for all of the changes in kinetic energy associated with the lowest order relativistic corrections.  

In \ref{8x3} the term $(\sqrt{\gamma_{jk} }\, d\xi)$ integrates to a magnitude of order, $1$, as decribed in Section 3. One can see that the size of the deviation is determined by the ratio of the interaction energy to the total relativistic energy. This deviation can be compared to the discrepancy associated with the conventional theory. The full relativistic kinetic energy can be represented as: 
$KE_{rel} \, = \, \sqrt{m_0^2c^4 +p^2c^2} \, - m_0c^2. $
Expanding this expression about $\mathbf{p}^2 \, = \, 0$, and retaining the two lowest order terms we get:
\begin{equation}\label{8x5} 
KE_{rel} \, = \, m_0c^2\{ [1 + \frac{1}{2}p^2 / m_0^2c^2 - \frac{1}{8}p^4/m_0^4c^4 + \, ... ] \, - 1 \} \, = \, \frac{1}{2}p^2 / m_0 - \frac{1}{8}p^4/m_0^3c^2 \, + \, ... 
\end{equation} 
Since the nonrelativistic formula for kinetic energy is 
$KE_{nr} \, = \, p^2 / (2m),$ the first-order correction can be rewritten as $- KE_{nr} * (\frac{1}{2})(KE_{nr} /mc^2).$ The change in kinetic energy is (approximately) equal to the change in potential energy. Thus, like the deviation implied by \ref{3x3}, the unaccounted for change in kinetic energy, 
$ \sim (\Delta \mathbf{\hat{ V}} / mc^2) * KE_{nr}$, in the conventional nonrelativistic theory is also proportional to the ratio of the interaction energy to the total relativistic energy.

It is also worth noting that for interactions involved in transitions from free states to free states the energy deviations that occur when the two systems are receding from one another tend to cancel those that occurred when they were approaching. 

For interactions that lead to transitions to or from bound states the conventional theory fails completely to account for energy changes. These situations involve the emission or absorption of photons with energies approximately equal to the change in kinetic energy. The standard theory is useful for calculating the energy discrepancies, but cannot explain them.

The deviations listed on the third line of \ref{8x2} involve two factors of the ratio of interaction energy to total relativistic energy. Thus, for typical nonrelativistic situations they are several orders of magnitude smaller than the those listed on the second line. There are a couple of compelling reasons for regarding these apparent deviations as artifacts of the nonrelativistic formulation rather than as real effects. First, 
they are less than radiative effects by a ratio of $(c/v)$. Second, in cases with much higher ratios of interaction energy to total relativistic energy, creation and annihilation processes would become relevant, and antiparticle content of the wave would need to be considered. Since there is always some small amount of antiparticle content in any localized wave function\footnote{For a calculation of the antiparticle content of localized wave packets see the discussion by Bjorken and Drell in \cite{Bjorken_Drell} (p.39). The ratio of the apparent deviation to radiated energy is based on the classical Larmor formula which yields a ratio of radiated power to energy change of 
$ (\frac{2}{3})\frac{c}{v} \mathbf{\big{ [}}\frac{(e^2/r)} {(mc^2)}\mathbf{\big{ ]}}^2. $ } it is reasonable to view these small discrepancies as resulting from the limitations of the nonrelativistic formulation.

To summarize, wave function collapse is not restricted to just the measured subsystem and the measurement apparatus. It also involves the systems with which the measured subsystem has previously interacted, even if the relevant entanglement is very small. The change in energy of the measured subsystem that is not attributable to interactions with the measurement apparatus is compensated for by correlated changes in these entangled systems. The small deviations that remain unaccounted for are attributable to the fact that the nonrelativistic theory is not able to describe certain forms of energy.

The next section examines the possibility of extending the current collapse proposal to account for relativistic effects.

\section{Possible Relativistic Extensions }
\label{sec:9}

Although the formulation of the collapse equation presented here is essentially nonrelativistic, as mentioned earlier the idea that wave function collapse is induced by the interactions that establish correlations between systems was originally motivated by the desire to reconcile relativity with the nonlocal aspects of quantum theory. The focus on these interactions stemmed from the central role that they play in the transmission of information. It is, therefore, reasonable to ask what are the prospects for extending this proposal to achieve full consistency with relativity. 

As argued in previous works\cite{Gillis_1,Gillis_2} full consistency requires a reconsideration of the foundations of \textit{both} quantum theory \textit{and} relativity. Recalling Einstein's repeated insistence that any inferences about the metric properties of space and time must be based on the observation of physical objects and processes\cite{Einstein_1,Einstein_2}, one must ask whether those properties are underdetermined by the probabilistic character of some physical processes. To what extent could such underdetermination hide some features of spacetime structure?

Consider that in quantum field theory Lorentz invariance is not guaranteed simply by the limiting speed of light. It is necessary to add the assumption that spacelike-separated interactions commute.\cite{Weinberg} Given the demonstration by Bell\cite{Bell_EPR} that quantum correlations imply some sort of nonlocal effects, what is it that prevents the assignment of an unambiguous sequencing to spacelike-separated interactions involving entangled systems? It is, precisely, the probabilistic character of nonlocal quantum effects. The reason that the relativistic description of spacetime is not disrupted by nonlocal quantum effects is that these effects are inherently probabilistic.

These considerations raise the possibility that there are spacetime features such as an evolving, spacelike hypersurface, along which the nonlocal quantum effects propagate. This structure remains hidden because of the nondeterministic nature of these effects. From this perspective the preferred reference frame used here to construct a low energy collapse theory can be viewed as just a simplified special case of such an evolving surface.

In order to extend the current proposal to a relativistic version the most straightforward approach would be to simply replace the preferred frame by a randomly evolving spacelike surface. Although any such surface does imply a sequencing of spacelike-separated interactions, since it evolves in a purely random fashion it remains impossible, in principle, to associate the specific sequencing with any observable physical effects. This approach also has the advantage that it maintains the equivalence of all reference frames, which is one of the essential features of relativity. Since the evolving surface would be unobservable in principle, the usual relativistic description of spacetime should emerge from the proposed extension of \ref{3x3}.

Such an extension would need to characterize interactions in a fully relativistic manner. So a stochastic operator based on interactions would involve more than scalar potentials. The account would need to incorporate all of the features of quantum field theory - massless particles, antiparticles, particle creation and annihilation, and other high energy effects. Such a development is, obviously, not trivial. But I believe that the challenges it would face are of a technical, rather than a fundamental conceptual nature.

\section{Summary}
\label{sec:10}

The hypothesis that wave function collapse is induced by the entangling interactions that generate decoherent branches leads to a stochastic collapse equation with several attractive properties. It  reduces to the Schr\"{o}dinger equation for stationary states and freely evolving systems, and deviates from it by a very small amount in situations that involve a few interacting elementary particles. It insures collapse to a definite outcome for systems of mesoscopic size with the correct probability and on a time scale that is consistent with our macroscopic experience. Because the strength and duration of the collapse effects are determined by the ratio of interaction energy to total relativistic energy it does not require the introduction of any new physical constants. 

Furthermore, it is consistent with exact conservation of momentum and orbital angular momentum in individual experiments, and it is consistent with energy conservation in those circumstances in which conventional nonrelativistic quantum theory correctly predicts such conservation. This consistency is based on the  recognition that all physical systems are quantum systems, and that they are almost always entangled to some extent. Since conserved quantities are shared across entangled branches the apparent nonconservation of a quantity in an elementary target system is compensated for by corresponding changes in (usually larger) systems with which it has interacted in the past.

Finally, compatibility with relativity can be achieved by replacing the fixed rest frame with an evolving spacelike surface. Nonlocal quantum effects propagate along the surface, and their nondeterministic nature prevents superluminal information transmission. This surface could be taken to evolve in a purely random fashion. This possibility respects the equivalence of all inertial frames, one of the central principles of relativity. With this approach one might reasonably hope to develop an account that explains measurement outcomes in terms of fundamental processes, while preserving the essential features of contemporary theory.


\begin{thebibliography}{99}
	

	
	
	\bibitem{Pearle_1976} Pearle, P.: Reduction of the state vector by a nonlinear Schr\"{o}dinger equation. Phys. Rev. D \textbf{13}, 857 (1976)
	
	\bibitem{Pearle_1979} Pearle, P.: Toward explaining why events occur. 
	Int. J. Theor. Phys. \textbf{18}, 489 (1979) 
	
	\bibitem{Gisin_1984} Gisin, N.: Quantum Measurements and Stochastic Processes.
	Phys. Rev. Lett. \textbf{52}, 1657 (1984).
	
	\bibitem{GRW} Ghirardi, G.C., Rimini, A., Weber, T.: Unified dynamics for 
	microscopic and macroscopic systems. Phys. Rev. \textbf{D34},  470-491  (1986) 
	
	\bibitem{Diosi_1} D\'{i}osi, L.: Continuous quantum measurement and the It\^{o} formalism. Phys. Lett. \textbf{129A}, 419 (1988) 
	
	\bibitem{Diosi_2} D\'{i}osi, L.: Quantum stochastic processes as models for state vector reduction. J. Phys. A \textbf{21}, 2885 (1988)
	
	\bibitem{Diosi_3} D\'{i}osi, L.: Localized solution of a simple nonlinear quantum Langevin equation. Phys. Lett. \textbf{132A}, 233 (1988).
	
	\bibitem{Gisin_c} Gisin, N.: Stochastic quantum dynamics and relativity.  Helv. Phys. Act. \textbf{62}, 363 (1989)
	
	\bibitem{GPR} Ghirardi, G.C., Pearle, P., Rimini, A.: Markov-processes in Hilbert-space and continuous spontaneous localization of systems of identical particles. Phys. Rev. A \textbf{42},  78  (1990)  


	\bibitem{Adler_Brun} Adler, S.L., Brun, T.A.: Generalized stochastic Schr\"{o}dinger equations for state vector collapse. J. Phys. A \textbf{34}, 4797 (2001) 


	
	\bibitem{Ghirardi_Bassi} Bassi, A., Ghirardi, G.C.: Dynamical Reduction Models. 
	Phys. Rep.  \textbf{379}, 257-427 (2003)
	
	\bibitem{Pearle_1} Pearle, P.: How stands collapse I. J. Phys. A: Math. Theor., \textbf{40}, 3189-3204 (2007)      	
	
	\bibitem{Brody_finite} Brody, D.C., Constantinou, I.C., Dear, J.D.C., Hughston, L.P.: Exactly solvable 	quantum state reduction models with time-dependent coupling. 	J. Phys. A: Math. Gen. \textbf{39}, 11029-11051. (2006)             
	
	\bibitem{Mertens_1}	Mertens, L., Wesseling, M., Vercauteren, N., Corrales-Salazar, A., van Wezel, J.: The inconsistency of linear dynamics and Born's rule. arXiv:2106.10136 [quant-ph]. (2021)

	
	\bibitem{Gillis_1}	Gillis, E.J.: Causality, Measurement, and Elementary Interactions. Found. Phys. \textbf{41}, 1757-1785 (2011); doi.org/10.1007/s10701-011-9576-x.



	\bibitem{Gillis_2}	Gillis, E.J.: Wave Function Collapse and the No-Superluminal-Signaling Principle. Int. J. Quant. Found. \textbf{5} 2 16-50 (2019). 


	\bibitem{Zeh} Zeh, H. D.: On the interpretation of measurement in quantum theory. Found. Phys. \textbf{1}, 69-76. (doi:10.1007/BF00708656) (1970)
	
	
	\bibitem{Zurek_ptr} Zurek,W.H.: Pointer basis of quantum apparatus: Into what mixture does the wave packet collapse? Phys. Rev. \textbf{D} \textbf{24}, 1516 (1981)
	
	\bibitem{Zurek_Darwin_1} Zurek, W.H.: Quantum Darwinism. Nat. Phys. \textbf{5}, 181-188. (doi:10.1038/nphys1202)  [quant-ph] (2009).
	
	
	\bibitem{Zurek_Darwin_2} Zurek, W.H.: Quantum Theory of the Classical: Quantum Jumps, Born's Rule, and Objective Classical Reality via Quantum Darwinism. Phil. Trans. R. Soc. \textbf{A 376}: 20180107. http://dx.doi.org/10.1098/rsta.2018.0107  (2018).
	

	\bibitem{Bell_fund} Bell, J.S.: Introductory Remarks. Phys. Rep. \textbf{137}, 7 (1986); 



	\bibitem{Dugic_1}  Dug\'{i}c, M.: What is ”System”: Some Decoherence-Theory Arguments. Int. J. Theor. Phys., \textbf{45}, 2215 (2006)

	\bibitem{Dugic_2}  J. Jekn\'{i}c-Dug\'{i}c, M. Arsenijev\'{i}c, M. Dug\'{i}c: Quantum Structures: A View of the Quantum World. (2013)


	\bibitem{Bertlmann_2} Thirring, W., Bertlmann, R.A., K\"{o}hler, P., Narnhofer, H.: Entanglement or separability: 	The choice of how to factorize the algebra of a density matrix. Eur. Phys. J. D \textbf{64}, 181 (2011)

	
	\bibitem{Bertlmann_1} Bertlmann, R.A.: Bell's Universe: A Personal Recollection. arXiv:1605.08081 [physics.hist-ph] (2016).


	\bibitem{Adler_Bassi}  Adler, S.L., Bassi, A.: Collapse models with non-white noises. J. Phys. A: Math. Theor. 40(50):15083. (2007)


	\bibitem{Adler_Vinante}  Adler, S.L., Andrea Vinante, A.: 
	Bulk heating effects as tests for collapse models 	Stephen L. Adler and Andrea Vinante
	Phys. Rev. A 97, 052119 – Published 18 May 2018
	DOI:https://doi.org/10.1103/Phys Rev A. 97.052119 
	

	\bibitem{Bassi_Ferialdi} Bassi, A., Ferialdi, L.: Non-Markovian dynamics for a free quantum particle subject to spontaneous collapse in space: General solution and main properties. Phys. Rev. A 80, 012116 (2009)
	https://doi.org/10.1103//PhysRevA.80.012116




	\bibitem{Mott} Mott, N.F.: The Wave Mechanics of $\alpha$-ray Tracks. Roy. Soc. Proc. \textbf{A126} 79 (1929) 





\bibitem{Bell_dBR_Bhm} Bell, J.S.: de Broglie-Bohm, delayed-choice, double-slit, and density matrix. Int. J. Quant. Chem. (\textbf{S14}):155 - 159 (1980); (doi.org/10.1002/qua.560180819) 	Reprinted in \textbf{Speakable and Unspeakable in Quantum Mechanics}, Revised edition (2004) pp. 111-116; ISBN-13: 978-0521523387 ISBN-10: 0521523389.



	\bibitem{Bell_imp_p_w} Bell, J.S.: On the impossible pilot wave. Found. Phys. \textbf{12}, 989-999 (1982); (doi.org/10.1007/BF01889272) Reprinted in \textbf{Speakable and Unspeakable in 	Quantum Mechanics}, Revised edition (2004) pp. 159-168; ISBN-13: 978-	0521523387 ISBN-10: 0521523389.


	\bibitem{Bohm_Hiley} Bohm, D., Hiley, B.J.: The Undivided Universe, Routledge, (ch. 7), New York (1993)


	\bibitem{Norsen_PW} Norsen, T.: The Pilot-Wave Perspective on Quantum Scattering and Tunneling. 	Am Journal of Physics \textbf{81}, 258 (2013); https://doi.org/10.1119/1.4792375


	\bibitem{Romano} Romano, D.: Bohmian Classical Limit in Bounded Regions. arXiv:1603.03060 [quant-ph]. (2016)


		
	\bibitem{Gemmer_Mahler} Gemmer, J.,  Mahler, G.: Entanglement and the  factorization-approximation. Eur. Phys. J. D  \textbf{17}: 385, (2001) https://doi.org/10.1007/s100530170012
		
		
	\bibitem{Durt_a} Durt, T.: Characterisation of an entanglement-free evolution.  arXiv:0109112  [quant-ph]  (2001).
		
	\bibitem{Durt_1} Durt, T.: Quantum Entanglement, Interaction, and the Classical Limit. Zeit. fur Nat. A \textbf{59}, 425 (2004).
		


	\bibitem{Gillis_3}	Gillis, E.J.: Interaction-Induced Wave Function Collapse Respects Conservation Laws.   arXiv:1803.02687 v6[quant-ph]. (2018)



	\bibitem{Marletto_Vedral}	Marletto, C., Vedral, V.: The Quantum Totalitarian Property and Exact Symmetries.   arXiv:2005.00138[quant-ph]. (2020)



	
	\bibitem{Bjorken_Drell}  Bjorken, J.D., Drell, S.D.: Relativistic Quantum Mechanics.  McGraw-Hill, New York (1964)





	\bibitem{Einstein_1} Einstein, A.: On the Electrodynamics of Moving Bodies. Annalen der Physik \textbf{17}, 891 (1905)



	\bibitem{Einstein_2} Einstein, A.: Geometry and Experience(address to the Prussian Academy of Sciences on January 27,1921). Reprinted in \textbf{Ideas and Opinions}, Dell Publishing 	Co.,Inc. New York (1981)


	\bibitem{Weinberg} Weinberg, S.: \textbf{The Quantum Theory of Fields I}, (p.198), Cambridge University Press, New York (1995)

	
	\bibitem{Bell_EPR} Bell, J.S.: On the Einstein Podolsky Rosen paradox. Physics \textbf{1}, 195-200 (1964); doi: 10.1103/PhysicsPhysiqueFizika.1.195. Reprinted in Speakable and Unspeakable in Quantum Mechanics, Revised edition (2004) pp. 14-21; ISBN-13: 978-0521523387 ISBN-10: 0521523389. 
	
	


\end{thebibliography}
\end{document}